\documentclass[twocolumn,showpacs,amssymb,nobibnotes,aps,floats,psfig,prb]{revtex4-1}
\usepackage{textcomp,amssymb,graphicx,epsf}
\usepackage{hyperref}
\usepackage{epstopdf}
\usepackage{amsbsy}
\usepackage{amsmath}
\usepackage{mathtools}
\usepackage {xcolor}
\usepackage{graphicx,epsfig}
\usepackage{float}
\usepackage[utf8]{inputenc}

\begin{document}
\title{Experimental and theoretical demonstration 
 of negative magnetization induced by particle size reduction in nano-form  Gd$_{1-x}$Ca$_x$MnO$_3$}
\author{Papri Dasgupta$^{1}$}
\author{Sanjukta Paul$^{^3}$}
\author{S. Kumar$^{1}$} 
\author{Sudhakar Yarlagadda$^{2}$} 
\author{Chandan Mazumdar$^{2}$}
\affiliation{$^{1}$Physics Division, Jadavpur University, Kolkata, 700032, India}
\affiliation{$^{2}$CMP Division, Saha Institute of Nuclear Physics, HBNI, Kolkata 700064, India}
\affiliation{$^{3}$School of Physical Sciences, National Institute of Science Education and Research, Jatni 752050, India}

\date{\today}
\begin{abstract}
 We report  a pervasive phenomenon of gradual emergence of negative magnetization in  typical 
$R_{1-x}D_x$MnO$_3$ orthorhombic perovskite manganites in nano form
(with $R^{3+}$ being a trivalent high-magnetic-moment 
ion from the heavier rare-earth elements  Gd, Tb, Dy  and $D^{2+}$
refers to divalent alkaline elements Sr, Ca, etc.), while   
 the bulk form 
manifests no negative magnetization.  
Extensive magnetization studies have been carried out in Gd$_{1-x}$Ca$_x$MnO$_3$ manganites 
around half doping.
We demonstrate experimentally  that particle size reduction  in nano-form manganites enhances the 
 propensity of the system to exhibit negative magnetization and provide a theoretical explanation for the phenomenon.
To test the universality of our findings, we have extended the measurement to Dy$_{0.6}$Ca$_{0.4}$MnO$_3$ and obtained similar results.
\end{abstract}

\maketitle
\noindent

\section{Introduction}
Over the past few decades, complex oxides such as the manganites have attracted intense research
interest. They display a fascinating tapestry 
of charge, spin, and orbital orders when a trivalent rare-earth element is replaced by a divalent alkaline element \cite{cnr,khomskii0,9,hotta,tokura,cheong,raveau}.
The rare-earth manganites $R{\rm MnO_3}$, for the compounds with
 rare-earth element $R$ = La--Dy, exhibit an orthorhombic
perovskite structure; on the other hand, compounds with $R$ = Ho--Lu form in hexagonal structure \cite{pauthenet,kimura}.

A large amount of research has been focused on
 features
of pure and doped $R{\rm MnO_3}$ when $R$ is a light rare-earth element with weak magnetic moment and large ionic radius.
On the other hand  insufficient attention has been paid to  pure and doped perovskite manganites involving heavy rare-earth elements (with large effective  moment 
and small ionic size  \cite{shannon}) such as $R{\rm =Gd,~Tb,~ and ~Dy}$  and divalent alkaline dopant elements Sr, Ca, etc. 
 Due to the usage of small-sized elements (${\rm Gd,~Tb,~ and ~Dy}$), a few important modifications have to be taken in account---the tolerance factor decreases 
 and the buckling and tilting of the
ideal cubic structure becomes pronounced. Additionally, interesting co-operative effects manifest due to
the large  magnetic moment of the ions in the rare-earth network---namely, a reversal
of the magnetization results due to the coupling of the two interacting magnetic networks \cite{yusuf,biswas}.
Furthermore,  these manganites, $R{\rm MnO_3}$ ($R{\rm =Gd,~Tb,~ and ~Dy}$), display multiferroicity with magnetic order being accompanied by electric polarization \cite{kimura2,tokura1,kimura3,cheong2,arima}.

Although permanent magnetic systems generally show positive magnetization values, there are some systems 
(having two sublattices) which may exhibit negative magnetization due to the negative-exchange coupling between 
the two ferromagnetically ordered sublattices 
\cite{yusuf}. 
However,  two individually ferromagnetic sublattices are not always  needed 
for the manifestation of negative magnetization.
There exists another interesting set of compounds, that also display negative magnetization,
wherein  a disordered paramagnetic sublattice is
coupled with an ordered ferromagnetic
subtlattce via negative-exchange interaction. 
{ In these compounds having two sublattices A and B, sublattice A orders  along the external field below 
a magnetic ordering temperature $T_{\rm C}$
and influences the sublattice B (which remains paramagnetic at the measuring 
temperature $T \ge T_{\rm C}$) to align opposite to the field of sublattice A. Below a certain temperature called the compensation temperature (which is sizeably smaller than $T_{\rm C}$),} if the moment of sublattice B 
becomes larger than that of A, the net magnetization becomes negative \cite{yusuf,biswas}. 
A finite magnetic anisotropy should also be present in
the system to prevent the rotation of the net
magnetic moment, below the compensation temperature, in the direction of
applied magnetic field.
Some examples of such 
compounds are orthorhombic perovskite manganites
$R_{1-x}D_x$MnO$_3$  
with $R^{3+}$ being  Gd, Tb, and Dy  and $D^{2+}$
referring to  Sr or Ca \cite{yusuf,biswas,4,5,snyder,ma,yingnan,nagaraja,jin,elizabeth}.

In recent years, research on compounds exhibiting negative magnetization has been gaining momentum
as these materials can be utilized in digital data storage systems
by exploiting their ability to reversibly switch between two distinct magnetization states (i.e., positive and negative) when  
the  temperature or an externally applied magnetic field or  light is varied  \cite{yusuf,7}.
 Bipolar switching of the magnetization can be employed to make magnetoelectronic devices such as volatile memories.
Furthermore, the low cost of synthesis of manganites by sol-gel technique makes them suitable for large-scale industrial application \cite{8}.

{{
{Generally, among the perovskite manganites, negative magnetization can be obtained in a low-bandwidth material  having 
mixed-valent Mn ions (i.e., Mn$^{3+}$ and Mn$^{4+}$).}} Low bandwidth  can be realized in systems containing the heavier
rare-earth elements (Gd--Dy) \cite{biswas}; mixed valency of Mn ions is obtained by partially substituting trivalent rare-earth ions 
by divalent alkaline-earth ions. These undoped heavier rare-earth manganites, including GdMnO$_3$ (whose doped version is studied here), have been
reported to exhibit antiferromagnetic (AFM) insulating behavior \cite{tokura1,4}.
We have chosen Gd-based manganites because of the simplicity resulting from the zero orbital angular momentum in  Gd$^{3+}$ ions.
With the substitution of divalent ions (e.g., Ca$^{2+}$), an equivalent number
of Mn$^{4+}$ ions are introduced in the Mn sublattice of Gd$_{1-x}$Ca$_x$MnO$_3$ to make the system charge neutral. 
With the Gd ions remaining paramagnetic down to  low temperatures, the Mn-Mn interaction leads to different
ordered states in Gd$_{1-x}$Ca$_x$MnO$_3$ depending on the relative percentage of Mn$^{3+}$ and Mn$^{4+}$ ions. 
The phase diagram of Gd$_{1-x}$Ca$_x$MnO$_3$, obtained from magnetoresistive measurements, report ferromagnetic insulating phase
for $x < 0.5$, which transforms into charge ordered (CO) state with AFM interactions for $x \ge 0.5$ \cite{4,10}. 
In the low-doped region ($0 < x < 0.5$), the ferromagnetically ordered Mn-spins force the Gd-spins to align opposite 
to its direction, i.e., opposite to $H_{\rm ext}$, due to a weak negative exchange interaction between them \cite{4}; thus, a ferrimagnet results
at very low temperatures. 
With decreasing temperature, 
as the contribution of the magnetic moment of paramagnetic Gd ions keeps on increasing
in the direction opposite to the external magnetic field, the net moment decreases and  can become negative. 
At higher doping concentrations ($x > 0.5$), 
the  AFM domains become larger in extent 
and percolate the sample
causing the total magnetization to increase steadily with decreasing temperature
\cite{10}.}

{{Although the phenomenon of negative magnetization   has been studied  in manganite compounds in  bulk form \cite{yusuf, 4, 10, 11}
and in thin-film form \cite{paturi2},
the role of  particle size in nano-form manganites
has not been investigated. In the nano systems surface effects in nanoparticles can alter the magnetic behavior.}}
{{In this work, 
we carried out a detailed study of Gd$_{1-x}$Ca$_x$MnO$_3$ compounds around half-doping (i.e., at x = 0.4, 0.5, and 0.6)  in their bulk and
nanoparticle
form (under the influence of external magnetic fields of various strengths) in order to understand the effect
of particle size on negative magnetization. 
While the bulk forms of Gd$_{1-x}$Ca$_x$MnO$_3$ (at $x=0.4~ {\rm and}~0.5$) exhibit charge ordering 
and no signature of negative magnetization, its 
nanoparticle forms 
do exhibit negative magnetization at low temperatures although they remain devoid of any charge ordering. However, when temperature is varied, both bulk and nano forms exhibit a hump in magnetization
for $x=0.4~ {\rm and}~0.5$; this is caused by ferromagnetic domains of Mn-spin sublattice aligning at higher temperatures followed by Gd spins in
this region aligning antiparallel to the Mn spins at lower temperatures due to a weak antiferromagnetic Mn-Gd coupling. 
Furthermore, in the antiferromagnetic domains of Mn-spin sublattice at any doping, 
 the localized Gd spins in this region (effectively decoupled from the antiferromagnetic Mn sublattice) produce a paramagnetic response to the weak external magnetic field. 
In nanoparticles, owing to the decrease in rigidity of the lattice
at the surface, the concentration of Gd ions (Ca ions)  is higher at the surface (core) because they produce lower (higher) tolerance factor.
Higher than the stoichiometric Gd concentration
at $x=0.4~ {\rm and}~0.5$ yields negative magnetization similar to Gd$_{1-x}$Ca$_x$MnO$_3$ at 
lower doping (i.e., lower  than $x=0.4~ {\rm and}~0.5$, respectively)  in the {{bulk/thin-film form}} \cite{4,paturi2}.
Additionally, the magnetization is also significantly enhanced due to double exchange in the outer shell of the nanoparticles.

The rest of the paper is organized as follows. In Sec. II,
we describe both the preparation methods, of bulk and nanoparticle 
samples of Gd$_{1-x}$Ca$_x$MnO$_3$, as well as the measurements performed.
Then, in Sec. III, we provide the theoretical framework employed to analyze the observed
magnetization and charge ordering. Next, in Sec. IV, we present our data showing the increase in propensity for 
negative magnetization as the particle size decreases in nano-form samples at various dopings.
We also offer our theoretical explanation for this phenomenon.
 We close in Sec. V with our
concluding observations and discuss possible future directions.

\section{Experimental details}

Polycrystalline samples of Gd$_{1-x}$Ca$_x$MnO$_3$ (x = 0.4, 0.5, 0.6) 
have been prepared by well known sol-gel technique. For the preparation of Gd$_{1-x}$Ca$_x$MnO$_3$,
Gd$_{2}$O$_3$,
CaCO$_3$ and MnO$_{2}$ were used as starting materials. Appropriate amount of oxides were separately dissolved in a HNO$_{3}$ solution (Oxalic acid is also added in case of MnO$_{2}$). These solutions were mixed, and an amount of citric acid equivalent to the total number of moles of metal ions was added under moderate heating and stirring conditions. Subsequently, mixture was slowly evaporated at 
{80$^{\rm o}$C--90$^{\rm o}$C} in water bath resulting in the formation of gel which was heated to 250$^{\rm o}$C to remove the organic matter and decompose the nitrates of the gel. The black ash thus obtained was ground and further heated to 550$^{\rm o}$C for 5 hours to kick-off the remaining organic matter and then pelletized. The pellets then finally sintered for 24 hours at 1250$^{o}$C to obtain bulk samples. Gd$_{1-x}$Ca$_x$MnO$_3$ 
nanoparticles of different sizes have been prepared by subsequently sintering the pellets at 700$^{\rm o}$C, 800$^{\rm o}$C, 900$^{\rm o}$C, 1000$^{\rm o}$C and 1100$^{\rm o}$C for 3 hours. In the rest of this paper, the corresponding nanoparticle forms are identified by mentioning the sintering temperature after the compound, for example Gd$_{0.5}$Ca$_{0.5}$MnO$_3$ annealed at 700$^{o}$C is written as Gd$_{1-x}$Ca$_x$MnO$_3$-700. The bulk samples are represented in the manuscript as Gd$_{1-x}$Ca$_x$MnO$_3$-bulk.
	X-ray diffraction (XRD) study was carried out at room temperature with TTRAX-III diffractometer (M/s Rigaku, Japan) using Cu-K$\alpha$ source having wavelength 1.5406 \AA. Lattice parameters have been estimated from XRD pattern using the FullProf software \cite{11a}.  Magnetization was measured using VSM-SQUID magnetometer (M/S Quantum Design Inc, USA). Temperature dependence of magnetization in zero-field-cooled (ZFC) and field-cooled (FC) protocol in the warming cycle has been carried out under different magnetic fields (100 Oe $<$ H $<$ 70,000 Oe) in the temperature range 5 K - 330 K. Field dependence of magnetization (M-H) and magnetic relaxation measurement, etc., has also been carried out.

\begin{figure}
[t]
\includegraphics[width=13.0cm]{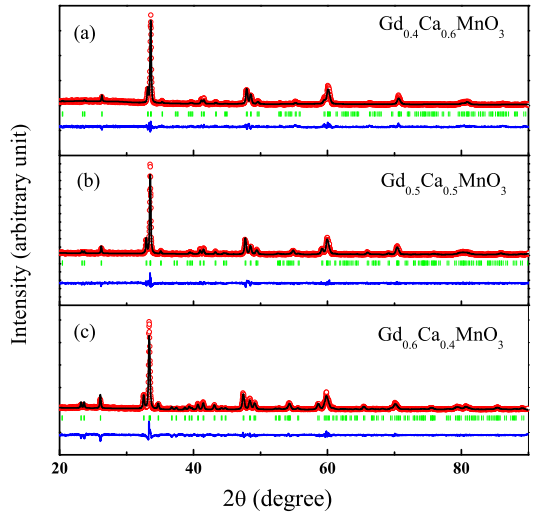}
\hspace{0.5cm}
\vspace{-2.5cm}
\caption{(colour online){
Room temperature X-ray diffraction pattern of bulk ${\rm Gd_{1-x}Ca_xMnO_3}$ (x = 0.4, 0.5, 0.6) compounds. The experimental data (red points), the calculated (black line) XRD patterns, allowed Bragg position by the space group (green lines), and the difference of experimental data and calculated values (blue lines) 
for all the compounds are displayed.
}}
\label{fig:1}
\end{figure}

Room temperature X-ray diffraction pattern of polycrystalline bulk samples of Gd$_{1-x}$Ca$_x$MnO$_3$ (x = 0.4, 0.5, 0.6), 
prepared by sol-gel technique, is shown in Fig. \ref{fig:1}. All the XRD lines confirm to {\em Pnma} space group suggesting the materials formed in single phase within the limit of resolution of our diffraction experiment.

\section{Theoretical framework}
We will analyze magnetization in ${\rm Gd_{1-x}Ca_x MnO_3}$ (GCMO) structures (bulk or nanomaterials) using
ZFC, FC, and isothermal hysteresis curves. We first present the following Hamiltonian, involving a single band for simplicity, without sacrificing
any essential physics:
\begin{eqnarray}
H = H_{\rm KE} + H_{\rm pol} + H_{\rm M} 
.
\label{H_tot}
\end{eqnarray}
The kinetic energy term $H_{\rm KE}  $ is given by
\begin{eqnarray}
 H_{\rm KE} = -te^{-\alpha E_p/\hbar\omega}\sum\limits_{\langle i,j \rangle} 
\left [{\cos \left ({\theta_{i,j}\over 2}\right ) {c^{\dagger}}_{i} c_{j}} + {\rm H.c.}\right ] ,
\label{H_ke}
\end{eqnarray}
where $t$ is the fermion hopping amplitude  that is attenuated by
the electron-phonon coupling  $ (E_p/\hbar\omega)^{1/2}$ with $E_p$ being the polaron energy, $\omega$ the optical phonon frequency, and $\alpha$ a constant of the order unity; $c_{j}$ is the $e_{\rm g}$ electron destruction operator; $\theta_{i,j}$ is the angle between 
  $S=3/2$ spins (originating from localized $t_{\rm 2g}$ spins) at ${\rm Mn}$ sites $i$ and $j$; and
$\cos (\theta_{i,j}/2) $ provides the modulation due to large Hund's coupling between the spin of the hopping $e_{\rm g}$ electron 
and the spins of the localized $t_{\rm 2g}$ electrons \cite{gennes,Izyumov}. Now, the small size of the ${\rm Gd^{3+}}$
ion (with the  size being smaller than the space created by the surrounding
${\rm MnO_6}$ octahedra, i.e., with the tolerance factor being small)  induces buckling of the ${\rm Mn-O-Mn}$ bonds  leading to a small hopping $t$;
since the electron-phonon coupling is strong, the kinetic energy is small. Then,
in the presence of disorder (such as cation disorder), even when the disorder is weak, this leads to localization (in fact, site localization)
of electrons \cite{hwang}.

The second term $H_{\rm pol}$ in  Eq. (\ref {H_tot}), is the polaronic Hamiltonian given by
\begin{eqnarray}
H_
{\rm pol} = -\sum_{j,\delta}
\left [ \beta E_p +\frac{\left [ t_{j,j+\delta}\cos(\theta_{j,j+\delta}/2)\right ]^2 }{2 \gamma E_p } \right ]
n_j(1-n_{j+\delta}), 
\nonumber \\
\label{H_pol}
\end{eqnarray}
where the first coefficient $\beta E_p$ ($\beta$ being of the order unity) is due to cooperative electron-phonon interaction
and portrays nearest-neighbor (NN) electron-electron repulsion due
to incompatible distortions of NN oxygen cages surrounding occupied ${\rm Mn}$ ions.
The value of $\beta$ depends on the nature of compatibility in the orbital order [see Ref. \onlinecite{ME}]. 
Next,
the second coefficient $\left [ t_{j,j+\delta}\cos(\theta_{j,j+\delta}/2)\right ]^2 /(2 \gamma E_p )$ results from second-order perturbation theory
and involves virtual processes 
 of hopping of a fermion from an ${\rm Mn }$ site to the
 NN ${\rm Mn}$ site and back and is present even when we consider the simpler
Holstein model \cite{sdadsy1}.
The factor $\gamma$ is of the order unity and increases with the increase in occupancy of
the neighbors of the ${\rm Mn}$ site $j+\delta$ (see Fig. \ref{fig:2}); this is because of the NN repulsion felt by the fermion when it is virtually
present at site $j+\delta$ in the intermediate state of the second-order process \cite{sp}.
Thus, {\em the second coefficient in Eq. (\ref{H_pol}) shows that not only is NN ${\rm Mn}$ site
occupation
discouraged, the next-nearest-neighbor ${\rm Mn}$ site occupation is also discouraged although to a weaker extent.}
Furthermore, although $n_j$ is the total number of electrons  in both the $e_{\rm g}$ orbitals at the ${\rm Mn}$ site $j$,
it can only take a maximum value of 1 due to strong on-site electron-electron repulsion
and strong Hund's coupling. It is important to note that, even though the electrons are localized, 
the electrons can virtually hop to the neighboring site and produce NN ferromagnetic spin alignment as well as nearest-neighbor, 
next-nearest-neighbor, and next-to-next-nearest-neighbor charge repulsion
through the second coefficient 
$\left [ t_{j,j+\delta}\cos(\theta_{j,j+\delta}/2)\right ]^2 /(2 \gamma E_p )$ 
which is much smaller than the first coefficient $\beta E_p$. Thus  the charge-order energy scale set by the first coefficient $\beta E_p$
is higher than the ferromagnetic spin-order energy scale set by the second coefficient 
$\left [ t_{j,j+\delta}\cos(\theta_{j,j+\delta}/2)\right ]^2 /(2 \gamma E_p )$. 
\begin{figure}
\includegraphics[width=0.8\linewidth,angle=0]{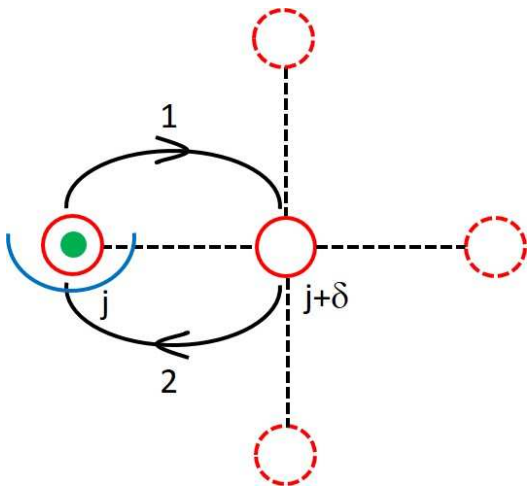}
\vspace{0.5cm}
\caption{(colour online) {Hopping process, corresponding to second-order
perturbation theory, shown for a two-dimensional Mn lattice. Schematic representation of a fermion,  originating at Mn site $j$,
hopping to its NN Mn site $j+\delta$ (the intermediate site) and
coming back. Here empty continuous-line circle corresponds to empty site, while
continuous-line circle with small dot indicates fermion position.  The intermediate site $j+\delta$ has each of its three NN Mn sites (depicted by dashed-line circles) either occupied by a fermion or empty. Semicircle 
at site $j$ represents full distortion of the lattice ions at that site with
corresponding energy $-E_p (+E_p)$ if the fermion is present
(absent) at that site. 
}}
\label{fig:2}
\end{figure}

The next term $H_{\rm M}$ in Eq. ({\ref{H_tot}) pertains
to the magnetic interactions;  in $\rm LaMnO_3$ it refers to the superexchange \cite{anderson} interactions which generate A-AFM order. 
In Gd$_{1-x}$Ca$_x$MnO$_3$, it is given by
\begin{eqnarray}
 H_{\rm M} =  \sum\limits_{\langle i,j \rangle} J_{i,j} \vec{S}_i \cdot \vec{S}_j 
 + J_{\rm Mn-Gd} \sum\limits_{\langle i, \eta \rangle} \vec{S}_i \cdot \vec{\cal{S}}_{\eta}  ,
 \label{H_se1}
\end{eqnarray}
where $J_{i,j}$ is the ${\rm Mn-Mn}$ superexchange coupling between the spins on the nearest neighbors 
and $J_{\rm Mn-Gd}$ ($< |J_{i,j}|$) is a weaker antiferromagnetic coupling
between a ${\rm Gd^{3+}}$ ion at the center of the unit cell (at location $\eta$) and its eight nearest-neighbor ${\rm Mn}$ ions at the corners of
the unit cell \cite{snyder}. 
In the above expression, the magnitude of the superexchange coefficient $J_{ i,j}$ is much smaller than
the  second coefficient 
$\left [ t_{j,j+\delta}\cos(\theta_{j,j+\delta}/2)\right ]^2 /(2 \gamma E_p )$ in Eq. (\ref{H_pol}).
At low temperatures,  the antiferromagnetic coupling $J_{\rm Mn-Gd}$ can produce a metastable state with  negative magnetization
(actually a ferrimagnetic state)
when at higher temperatures  a ferromagnetic state results. 
The coupling between the rare-earth network and the manganese network is a feature absent in  La$_{ 1-x}$Ca$_{ x}$ MnO$_3$.

\section{Results and discussions}
\begin{figure}
[t]
\includegraphics [width=1.9\linewidth,angle=0]
{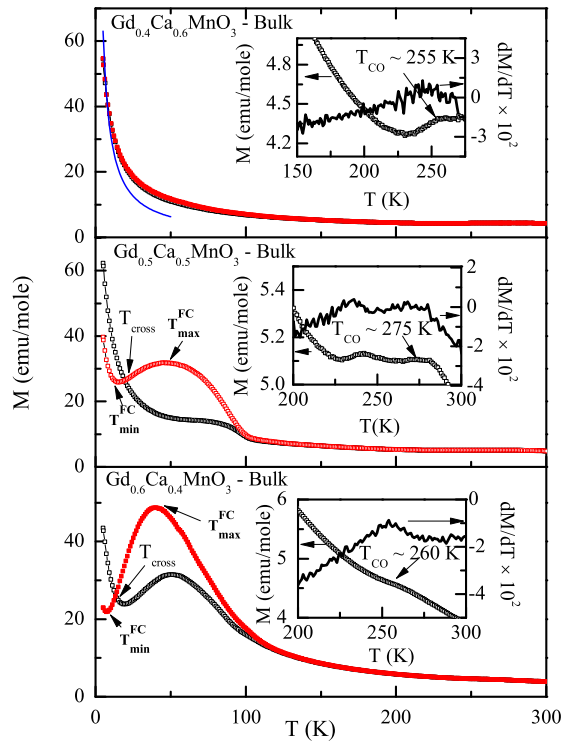}
\vspace{-2.5cm}
\caption{(colour online) {Temperature dependent magnetization of Gd$_{1-x}$Ca$_x$MnO$_3$-bulk ($x$ = 0.4, 0.5, 0.6) at H = 100 Oe. FC (ZFC)
data is plotted in red (black). Inset shows the enlarged view of the magnetization curve (left axis) to point out the CO temperature  and the corresponding peak in the derivative of magnetization curve (right axis). For $x = 0.6$ case, theoretical response for free Gd$^{3+}$ ions  (plotted in blue below 50 K) shows their  dominant contribution to the  total magnetization at lower temperatures. 
}} 
\label{fig:3}
\end{figure}

\begin{figure}
[b]
\includegraphics [width=1.4\linewidth,angle=0]
{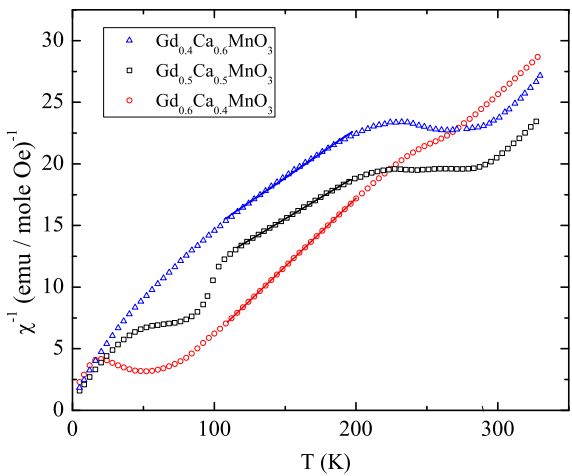}
\vspace{-2.6cm}
\caption{(colour online) {Inverse magnetic susceptibility of bulk Gd$_{1-x}$Ca$_x$MnO$_3$ ($x$ = 0.4, 0.5, 0.6) measured at 100 Oe
under ZFC conditions. 
The linear regions,  indicating Curie-Weiss behavior, are depicted by solid lines. In the depicted linear regions,  ZFC and FC curves coincide
as shown  in Fig.~\ref{fig:3}.
}} 
\label{fig:chi-1}
\end{figure}
In this paper we analyze Gd$_{1-x}$Ca$_x$MnO$_3$ samples at concentrations in the neighborhood of half-doping, i.e., for $x =0.4$, $0.5$, and $0.6$.
Now, based on the phase diagram for bulk samples 
{(see Ref. \onlinecite{10})}, for $x< 0.5$ we have a ferromagnetic
insulator, whereas for $0.5 \le x \le 0.8$ we get an antiferromagnetic insulator at lower temperatures.
{For $x< 0.5$, based on Ref. \onlinecite{sp}, ferromagnetic insulator can be explained  as follows.}}
Since cooperative electron-phonon interaction is strong, 
    a NN electron-hole pair has a ferromagnetic interaction 
   $\left [ t_{j,j+\delta}\cos(\theta_{j,j+\delta}/2)\right ]^2 /(2 \gamma E_p )$ [as shown in Eq. (\ref{H_pol})]
   which is much stronger than other magnetic interactions ($|J_{i,j}|$ and $J_{\rm Mn-Gd}$).
   Hence, a  ferromagnetic cluster (magnetic polaron) is generated in the vicinity of a hole.
   In fact, the hole (through virtual hopping) will polarize electrons that are nearest-neighbor as well as those that are next-nearest-neighbor 
   and next-to-nearest-nearest-neighbor to form a magnetic polaron.
    A collection of interacting magnetic polarons will create a ferromagnetic-insulator region. 
    As regards the doping region  $ x \ge 0.5 $, several theories have been put forth to explain antiferromagnetism \cite{khomskii0,hotta};
    we do not propose any new explanation beyond these theories for antiferromagnetism.

   We will now provide an understanding of the magnetization displayed in
   the ZFC, FC, and hysteresis curves for bulk and nanoparticle forms.
   
 \subsection{FC and ZFC cases for weak external field}  
 \subsubsection{Bulk samples}
   We will first analyze the temperature dependent magnetization M(T) curves for Gd$_{1-x}$Ca$_x$MnO$_3$-bulk ($x$ = 0.4, 0.5, and 0.6), 
     obtained under ZFC and FC conditions at 100 Oe external field, shown in Fig.~\ref{fig:3}.
       Enlarged view of the M(T) curves below 300 K, shown in insets of Fig.~\ref{fig:3}, displayed discernible humps in the 
     temperature range ~ 255 K--275 K. The anomaly can, however, be prominently manifested on taking the derivative of the M(T) curves (see insets and their right axes  in Fig.~\ref{fig:3}). This is a commonly observed behavior in many manganite systems including Gd$_{1-x}$Ca$_x$MnO$_3$ 
     ($x \geq 0.5$) and has been explained to have arisen due to CO at temperature ~ T$_{\rm CO}$ \cite{13}. It may be noted here that the signature of CO in our samples has been extended even beyond the reported region of $x \geq 0.5$  
          (see Ref. \onlinecite{10}), as the same signature is observed in our Gd$_{0.6}$Ca$_{0.4}$MnO$_3$ sample [see Fig. ~\ref{fig:3} bottom (inset)]. 
          Below T$_{\rm CO}$, in the temperature range 110 K--200 K, inverse susceptibility  [$\chi^{-1}$({\rm T})] curves  of all these compounds 
     (as shown in Fig.~\ref{fig:chi-1}) follow Curie-Weiss (CW) behavior, i.e.,  a linear behavior with $\chi^{-1}({ T})=(T-\theta_{\rm CW})/C$.
     For Gd$_{0.6}$Ca$_{0.4}$MnO$_3$, the values of $\theta_{\rm{CW}}$ and $C$, as estimated from the CW behavior, are  44.7 K and  9.02 emu K/mole,
     respectively.
     The positive value of $\theta_{\rm{CW}}$ indicates that the dominant exchange interaction is of the ferromagnetic type.
     However,  indicating antiferromagnetism, $\theta_{\rm CW}$ is negative for Gd$_{0.5}$Ca$_{0.5}$MnO$_3$ and Gd$_{0.4}$Ca$_{0.6}$MnO$_3$ 
     (as can be seen in  Fig.~\ref{fig:chi-1}); $\theta_{\rm{CW}}=-73.6$ K and $C = 14.49$ emu K/mole for Gd$_{0.5}$Ca$_{0.5}$MnO$_3$ 
     and $\theta_{\rm{CW}}=-103.5$ K and $C = 9.00$ emu K/mole for Gd$_{0.4}$Ca$_{0.6}$MnO$_3$.  We note that the magnetic behaviors in our samples 
     of $x =0.4$, $0.5$, and $0.6$ are in agreement with those reported in Ref. \onlinecite{10}.

   Now, the system at $x=0.4$, based on the phase diagram in 
      {Ref. \onlinecite{10}} and Fig.~\ref{fig:chi-1}, is primarily ferromagnetic with small domains that are
   antiferromagnetic. 
      {Existence of small antiferromagnetic domains at $x=0.4$ is expected due to the doping proximity to $x=0.5$ where the system is antiferromagnetic; {
      {this will be further justified/clarified below.}} Hence, for the FC case,  at temperatures ${\rm T > T_{\rm Mn-Gd}}$,
   pertaining to the thermal energy larger than the magnetic energy 
      for the weak antiferromagnetic coupling $J_{\rm Mn-Gd}$, only the ferromagnetic fields
   (generated by the magnetic polarons) are
   relevant; here, ferromagnetism decreases with increasing temperatures. At temperatures below
   ${\rm T_{\rm Mn-Gd}}$, the antiferromagnetic coupling $J_{\rm Mn-Gd}$ starts becoming relevant
   and the  ${\rm Gd}$ spins (with $S=7/2$) in the percolating ferromagnetic cluster align antiparallel to the percolating  cluster.
   At temperatures ${\rm T < T_{\rm Mn-Gd}}$, the
      ${\rm Gd}$ spins that are aligned opposite to the percolating ferromagnetic cluster
   start lowering the overall magnetism.
    {Contrastingly}, the ${\rm Gd}$ spins  in the small antiferromagnetic domains
   start aligning (as temperature is lowered, i.e., for $T \lesssim 250$ K) due to the external weak magnetic field (H = 100 Oe) and contribute to increasing magnetism. This is because, 
   {{\em  in the small antiferromagnetic domains, there is no net coupling between ${\rm Gd}$ spins
   and the antiferromagnetic lattice of  ${\rm Mn}$ spins.} Needless to say that, when the ${\rm Mn}$
    spins at the 8 vertices of a unit cell are aligned antiferromagnetically, the total spin of the ${\rm Mn}$ spins at the 8 vertices  is zero; 
    hence, when a ${\rm Gd}$ spin (with $S=7/2$) is introduced at the center of the cube, 
    the central ${\rm Gd}$ spin has no net coupling
    to the total spin of the 8 vertices.}
      
   Now, the ZFC curve gives a picture that is qualitatively similar to that of the FC curve.
   However, the ZFC curve is further away from equilibrium compared to the FC curve because its ferromagnetic
   domains (compared to the FC case) are less aligned with the external field. This is because  the starting state for the warming cycle of the ZFC case
     was obtained by cooling at zero field. Thus, the magnetization peak  of the ZFC curve is lower
     than that of the FC curve. 
          {Consequently, there is a pronounced
     ZFC-FC bifurcation.}

     Next, at $x=0.5$, based on the phase diagram in 
          {Ref. \onlinecite{10}, the system
     is primarily antiferromagnetic with small clusters that are ferromagnetic. 
          {In the FC case, 
     the ferromagnetic clusters are strongly aligned with the external field and contribute to the magnetism
     similar to the larger ferromagnetic regions of the $x =0.4$ case. Again, for the ZFC case, since the
     small ferromagnetic clusters are less aligned with the external field (than in the FC case), we get a smaller magnetization. 
     Since the ferromagnetic domains are less dominant for the case of $x=0.5$ compared to that of $x=0.4$,
     the ZFC-FC bifurcation is less pronounced and the humps are also smaller for $x=0.5$.}
     Below    ${\rm T \lesssim 250}$ K, the ${\rm Gd}$ spins  in the large antiferromagnetic domain
   start aligning due to the external weak magnetic field (H = 100 Oe) and contribute to larger magnetism
   compared to $x=0.4$ case. 
   
      {We also notice that the ZFC and the FC curves cross between the two magnetization extrema (i.e., the maximum and the minimum)
   for x = 0.4 and 0.5.
   The ZFC-FC crossover is obtained at  T$_{\rm cross}$ below which the FC M(T) curve attains values lower than the ZFC M(T) curve. This is
      because the ZFC curves have larger AFM domains and smaller ferromagnetic domains (that are less aligned with 
      the direction of H$_{ext}$) than the 
            FC case. Furthermore, in the ferromagnetic domains at $x=0.4$, the negative magnetization  due to the ${\rm Gd}$ spins is more than the positive
     magnetization due to the ${\rm Mn}$ spins.}
        {In fact, based on the 
    formula  Gd$_{1-x}$Ca$_x$MnO$_3$, when ${\rm Gd}$ spins are antiparallel to the ${\rm Mn}$ spins and the weak external magnetic field,
    we can work out the effective spin of a unit cell to be $-\frac{7(1-x)}{2}+2(1-x)+\frac{3x }{2}=-1.5+3x$;
    here, the axis of quantization is in the direction of the weak external field. Thus, at very low temperatures,   magnetization 
    in a cluster (containing ferromagnetic Mn sublattice) can be negative only when  $-1.5+3x$ is negative, i.e.,  $x < 0.5$.} 
    In bulk samples at $x=0.4$, as depicted in Fig.~\ref{fig:3}, it is important to note that  the magnetism due to Gd spins in antiferromagnetic clusters wins over the negative
    magnetization contribution from the ferromagnetic domains and keeps the magnetism positive, i.e, aligned with the external magnetic field.

   Lastly, at $x=0.6$ the system is essentially fully antiferromagnetic.
   Thus, for both ZFC and FC cases, at lower temperatures only the  ${\rm Gd}$ spins  get
   more aligned  due to the external weak magnetic field and magnetization increases in a similar fashion.
      {Furthermore, as shown in Fig.~\ref{fig:3}, neither a ZFC-FC bifurcation nor a hump are  visible due to negligible amount of ferromagnetic domains.
   Also plotted (in blue) is the paramagnetic response for free Gd spins $M= CH/T$ where $C=Ng^2 S(S+1)\mu_B^2/(3k_B)$ with  $S=7/2$ and $N$ being
   the number of Gd ions per mole;
   it is clearly demonstrated that paramagnetic Gd ions in the AFM region give the dominant contribution at lower temperatures.}
      {As doping $x$ increases,  ${\rm Gd}$ concentration decreases resulting in smaller contribution to magnetization 
   by the paramagnetic ${\rm Gd}$ spins at lower temperatures (compare magnetization values for $x=0.6$ with those for $x=0.5$ at
   ${\rm T < T^{FC}_{min}}$ in Fig.~\ref{fig:3}).}}

  {It is also important to note that, since the charge-order energy scale (set by $\beta E_p$ in Eq. (\ref{H_pol})) is higher than both the ferromagnetic spin-order energy scale (set by $\left [ t_{j,j+\delta}\cos(\theta_{j,j+\delta}/2)\right ]^2 /(2 \gamma E_p )$ 
   in Eq. (\ref{H_pol})) as well as the AFM energy scale $J_{ij}$ (mentioned in Eq. (\ref{H_se1})),
   the charge ordering occurs at a much higher temperature than the magnetic ordering (as shown in Fig.~\ref{fig:3}). }
   
\begin{figure}
[b]
\includegraphics [width=1.9\linewidth,angle=0]
{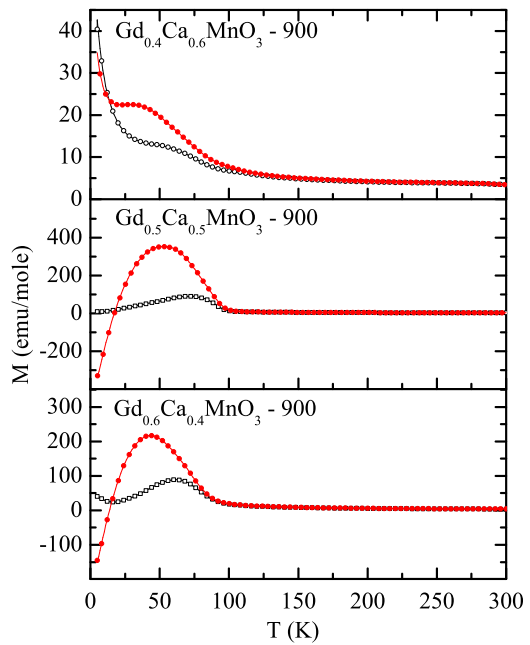}
\vspace{-3.0cm}
\caption{(colour online) {Temperature dependent magnetization curves of Gd$_{1-x}$Ca$_x$MnO$_3$  - 900 ($x$ = 0.4, 0.5, 0.6) nanoparticles under a field of 100 Oe. FC (ZFC)
data is depicted in red (black). The continuous lines are meant to be guides to the eye.}}
\label{fig:4} 
\end{figure}

\subsubsection{Nano-form samples}
   Next, we will  analyze the magnetization curves  for manganites in nanoparticle forms at the same dopings
   $x=0.4$, $0.5$ and $0.6$ as was done for the manganites in bulk form. 
      {We did not observe any charge order 
      for the nano forms of GCMO reported in this paper; this is in contrast to the bulk case where we witnessed CO (see Fig.~\ref{fig:3}). }
      It is interesting to note that, for the lower-doped cases among the nano forms considered, i.e., for  $x=0.4$ and $0.5$  as shown in  Fig. \ref{fig:4},
   we get negative magnetization
   at lower temperatures (i.e., $T \lesssim 20~ {\rm K})$ for the FC case. In the nanoparticles, the unit cells are less
   rigid at the surface of the nanoparticles compared to the center. Consequently, it is natural
   that unit cells with smaller tolerance factors  
      (which lead to distortion effects such as more buckling of ${\rm Mn-O-Mn}$ bonds)
      can be accommodated better at the surface than at
   the center. Hence, since ${\rm Gd}$ ion has a smaller radius than the ${\rm Ca}$ ion, the
   concentration at the surface (center) of the nanoparticle for the ${\rm Gd}$ ions is higher (lower)
   than the overall stoichiometric concentration  given by the formula  Gd$_{1-x}$Ca$_x$MnO$_3$.
   Specifically, at $x=0.4$ ($0.5$), the outer shell has less than $0.4$ ($0.5$) holes per unit cell.
   It is important to realize that only when the effective concentration of ${\rm Gd}$ ions is more than
   $0.5$ in a region can we get negative magnetization in that region; this is because, then the average effective spin of  ${\rm Gd}$ ions is $> 7/4$
   and the average effective spins of ${\rm Mn}$ ions $< 7/4$ and the spins of ${\rm Gd}$ ions align antiparallel to those  of ${\rm Mn}$ ions which
   are in the direction of the weak external magnetic field (H = 100 Oe).
    Furthermore, there is greater charge delocalization at the surface because charge order is weaker here. Consequently,
   the charge polarizes beyond the nearest neighbor unlike the bulk situation modeled by Eq. (\ref{H_pol});
     the manganite in the nanoparticle form is more ferromagnetic than in the bulk form at the same effective concentration
   because of stronger double exchange phenomenon. 
     For these reasons, compared to the bulk form, we have larger magnetism in nanoparticle form
   at higher temperatures and also larger negative magnetization at low temperatures. Accordingly, we see that
   the FC curves show large negative magnetization at low temperatures with the uncoupled ${\rm Gd}$ spins in the
   antiferromagnetic core
   making only a weak positive magnetization contribution at low temperatures. 
      {Thus the observed magnetism is primarily from the  outer shell
   for the cases  $x=0.4$ and $0.5$. Furthermore, for the above mentioned
   reasons, the FC curve for the $x=0.6$ nano-system is similar to that of the bulk ${\rm Gd_{0.5}Ca_{0.5} MnO_3}$ system
   with the peak magnetization being much smaller than at  $x=0.4$ and $0.5$.}
    
  {As regards the ZFC curves  at $x=0.4$, $0.5$ and $0.6$, since all the ferromagnetic domains 
  are not fully aligned with the external field, we observe a weaker effect due to the outer shell.
  Unlike the FC curves, the ZFC curves are 
    non-negative due to the fact that ferromagnetism is weaker
  when cooled in the absence of the external field; at lower temperatures, the negative magnetization resulting
  from the ferromagnetic outer shell is weaker than the positive magnetization resulting from  the uncoupled ${\rm Gd}$ spins in the
   antiferromagnetic core.}

We now note that as particle size increases in various nano forms of GCMO, the FC  (ZFC) curves gradually tend toward
   the FC (ZFC) curves  of the bulk system as expected. In confirmation, as particle size increases in Fig. \ref{fig:5} for ${\rm Gd_{0.5}Ca_{0.5} MnO_3}$, the size of the hump decreases and the negative magnetization
   at lower temperatures  reduces.

   \begin{figure}
[t]
\includegraphics[width=0.95\linewidth,angle=0]{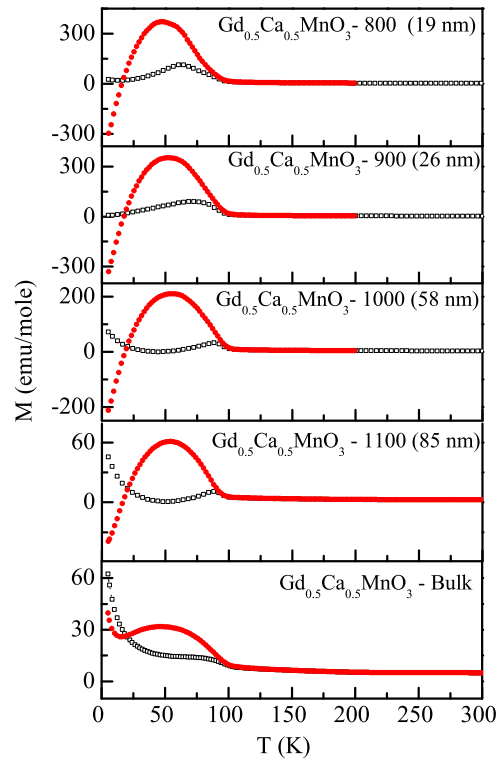}
\vspace{-2.0cm}
\caption{(colour online) {Temperature dependent magnetization curves of Gd$_{0.5}$Ca$_{0.5}$MnO$_3$, under a field of 100 Oe,
for different 
annealing temperatures leading to different nanoparticle forms. The FC (ZFC) data is presented in red (black).}}
\label{fig:5}
\end{figure}

  We will now make the important observation that negative magnetization at dopings $x=0.4$ and $0.5$ in nano-form GCMO is a metastable
  state because net magnetization is opposite to the applied field. Also, we expect ZFC curves to be further away
  from equilibrium than the FC curves because exposing the samples to magnetic field (while
  cooling) in the FC case helps them tend towards equilibrium. It is of interest to note that by-and-large
  the magnetization curves for the FC case (shown in Figs. \ref{fig:3} and \ref{fig:4}) and the remnant magnetization ${\rm M_R}$ curves (depicted in Fig. \ref{fig:MR}) are similar at higher temperatures (i.e., $T > 20$ K).
  Thus, it is reasonable to expect that the FC curves and the remnant magnetization curves (obtained under
  very different conditions) are representative
  of the equilibrium physics at these higher temperatures  (i.e., $T > 20$ K).

  \begin{figure}
[t]
\includegraphics [width=1.3\linewidth,angle=0]
{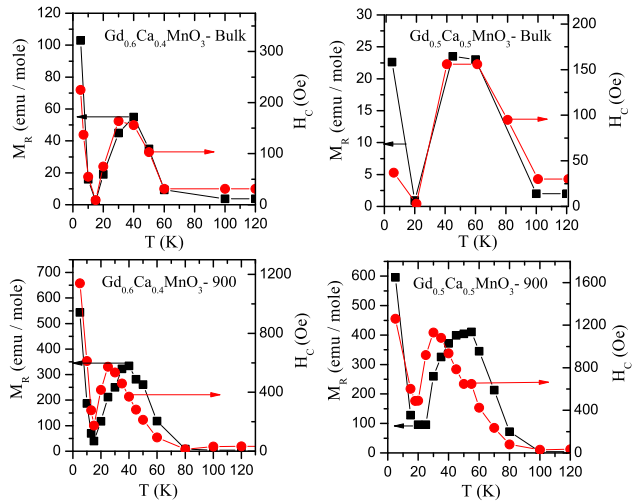}
\vspace{-2.0cm}
\caption{(colour online) {Temperature dependence of remnant
magnetization ${\rm M_R}$ and coercive field ${\rm H_C}$ obtained from magnetic hysterisis loops.}}
\label{fig:MR} 
\end{figure} 
  
  \subsection{Remnant magnetization analysis at low temperatures}
       {At lower temperatures
   (i.e., $T \lesssim 20~ {\rm K})$, we will analyze the remnant magnetization curves for the nanoparticle forms and the bulk forms separately.
   The expectation at equilibrium is that the ${\rm Gd}$ spins in the ferromagnetic regions
     will align with the external field  whereas the ${\rm Mn}$ spins in the same 
     regions will be antiparallel due to their antiferromagnetic coupling to
    the ${\rm Gd}$ spins.
   At 5 K, although
    the nanoparticle forms in Fig. \ref{fig:MR} display positive magnetization, they show much larger magnetization than when ${\rm Gd}$ spins
    are parallel to  the external field  and the ${\rm Mn}$ spins are antiparallel, i.e., much larger magnetization than obtained
    by reversing the sign of the
   magnetization in the nano-form curves for $x=0.4$ and $0.5$  in  Fig. \ref{fig:4} at 5 K. 
   Now, it is to be noted that during the isothermal magnetization measurements, due to exposure to high magnetic fields,
   the ${\rm Gd}$ spins in the ferromagnetic regions will be aligned with the external field. Consequently,
   we infer that the ${\rm Mn}$ spins are not aligned antiparallel
   to the external magnetic field.
   Hence, the  remnant magnetization curves also do not correspond to equilibrium.}
   
     {As regards the bulk forms at $T \lesssim 20~ {\rm K}$, for $x=0.5$, 
  as explained earlier, the magnetization due to  ${\rm Gd}$ and ${\rm Mn}$ spins cancel each other
   in the ferromagnetic regions at low temperatures; the only contribution to magnetization is from the effectively uncoupled ${\rm Gd}$ spins in the antiferromagnetic region. Thus, the equilibrium value of magnetization for $x=0.5$ is expected to be similar to the FC value in Fig. \ref{fig:3} which is slightly
   higher than the remnant magnetization in Fig. \ref{fig:MR}. 
   
   Next, the bulk form at $x=0.4$, in a unit cell, has an effective spin 
   for the ${\rm Gd}$ ions $\big ( =\frac{7(1-x)}{2} =2.1\big )$  which is larger than the effective spin for the  ${\rm Mn}$ ions $\big (=2(1-x)+\frac{3x }{2}=1.8 \big )$.
   Furthermore, the magnetization due to ${\rm Gd}$ ions in the ferromagnetic (antiferromagnetic) regions ${\rm M_{Gd}^{F}} $ (${\rm M_{Gd}^{AF}} $)
   and the magnetization due to the ${\rm Mn}$ ions in the ferromagnetic  regions ${\rm M_{Mn}^{F}} $, together contribute as 
   $-{\rm M_{Gd}^{F}}+ {\rm M_{Mn}^{F}}+ {\rm M_{Gd}^{AF}} $ at 5 K to the total magnetization ${\rm M_{tot}}$ in Fig. \ref{fig:3}.
   Then, based on the FC curve for $x=0.4$ in Fig. \ref{fig:3}, 
      we see that at 5 K ${\rm M_{tot}} \sim 20  $ emu/mole, ${\rm M_{Mn}^{F}} \sim 50$ emu/mole (based on FC curve value at 50 K), ${\rm M_{Gd}^{F}} \sim 50 \times 2.1/1.8 \sim 60 $ emu/mole;
    consequently,  at 5 K ${\rm M_{Gd}^{AF}} = {\rm M_{tot}}-{\rm M_{Mn}^{F}}+{\rm M_{Gd}^{F}} \sim 30  $ emu/mole. 
    For the bulk-form case of $x=0.4$ in Fig. \ref{fig:MR}, where  the remnant
    magnetization value is  $\sim 100$ emu/mole at 5 K, it appears that the ${\rm Gd}$ spins in the ferromagnetic and antiferromagnetic regions,
    when aligned parallel to the external magnetic field, contribute as ${\rm M_{Gd}^{F}}+{\rm M_{Gd}^{AF}}\sim 90$ emu/mole. This implies
    that the contribution of the ${\rm Mn}$ spins  is $\sim 10$ emu/mole  which is certainly not an equilibrium situation. 
    At equilibrium, the magnetization contribution of the ${\rm Mn}$ spins should be negative (i.e., $\sim -50$ emu/mole). }

   It should also be noted that, as expected,  the variation of the coercive field  as a function of the temperature in Fig. \ref{fig:MR} follows a trend that is similar to that
   of the remnant magnetization. Various details of the hysterisis loops are given in Appendix A.

 \begin{figure}
[t]
\includegraphics[width=2.3\linewidth,angle=0]{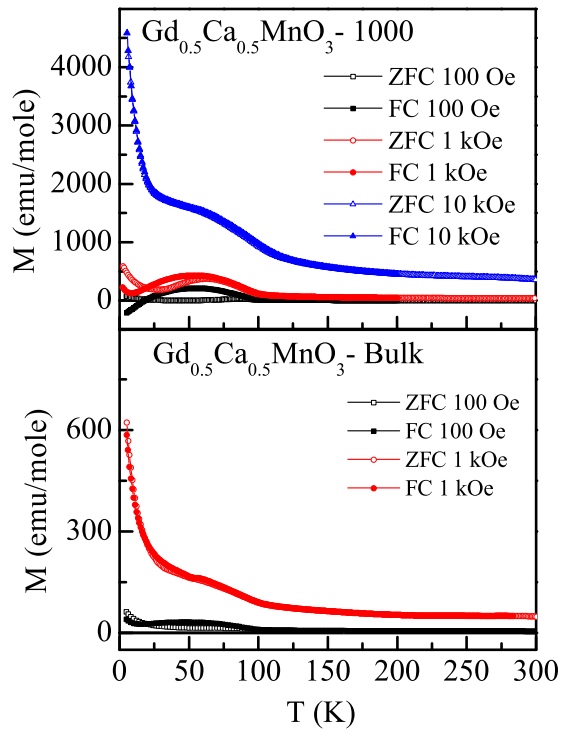}
\vspace{-4.5cm}
\caption{(colour online) {Effect of different applied fields on the temperature dependent magnetization curves of 
{
{Gd$_{0.5}$Ca$_{0.5}$MnO$_3$\,-\,Bulk }} and Gd$_{0.5}$Ca$_{0.5}$MnO$_3$\,-\,1000 systems.}}
\label{fig:7} 
\end{figure}

   \subsection{FC and ZFC cases for stronger external fields}
   
   Lastly, we study the effect of stronger magnetic fields (i.e., fields  stronger than 100 Oe) on the temperature 
   dependence of magnetization in both bulk and in nanoparticle forms.
   In Fig. \ref{fig:7}, we note that, as expected, at higher magnetic fields both the FC and ZFC
   curves for bulk and nano forms of GCMO samples increase in magnetization and the curves shift upwards. 
   We also see that, as applied
   magnetic field becomes stronger,  in  the lower-doped among the nano compounds considered (such as ${\rm Gd_{0.5}Ca_{0.5} MnO_3}$), 
   the system tends to attain  equilibrium state (with the Gd spins in the
ferromagnetic regions tending to align with the external field) thereby
   becoming more prone to positive magnetization at low temperatures. Additionally, as the magnetic field increases,
   the degree of ZFC-FC bifurcation decreases; at sufficiently high fields, the FC and ZFC curves merge. The effect of increasing  magnetic fields on the magnetization for bulk form
   of GCMO at different dopings is shown in Appendix B.
   Furthermore,
   as the particle size increases in the lower-doped ones among the GCMO systems considered (such as ${\rm Gd_{0.5}Ca_{0.5} MnO_3}$), 
   there is a decrease in the magnetic field needed to produce positive magnetization and make the systems attain equilibrium
    at low temperatures, i.e., at $T \sim 5~{\rm K}$
   (see Fig. \ref{fig:7}  and Fig. \ref{fig:8} in Appendix C).

\section{Conclusion and future direction}
{In this work, by extending the framework  developed earlier for ferromagnetic insulators \cite{sp}
to ferrimagnetic insulators, we explain negative magnetization.
Our picture of negative magnetization, invoking surface effects of nanoparticles, is comprehensive and is valid not only at 
the doping levels considered in this work (i.e., $x \ge 0.4$)
but also at dopings lower than those considered.
We also provide an understanding of the size effect of the particles on negative magnetization in  GCMO of various forms 
(i.e., bulk, thin-film and nano forms).
}
 
In future, experiments to probe
the local concentration of Gd ions are needed to further validate our picture that, near the surface of nanoparticles, the concentration of Gd
ions is higher  than that indicated by the chemical fomula Gd$_{1-x}$Ca$_x$MnO$_3$.
Making thin films and using cross-sectional transmission electron microscopy,  to measure the surface Gd-ion concentration as a function of depth,
is likely to clarify the situation.}

\begin{figure}
[t]
\includegraphics[width=2.0\linewidth,angle=0]{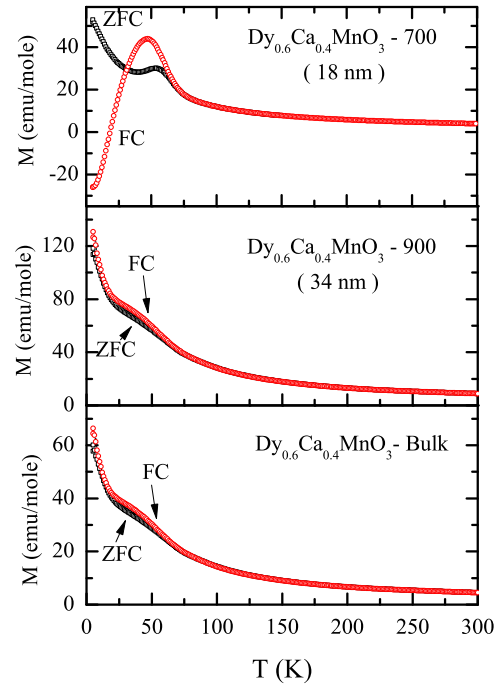}
\vspace{-3.0cm}
\caption{(colour online) {Temperature dependent magnetization curves of Dy$_{0.6}$Ca$_{0.4}$MnO$_3$ for different particle sizes and annealing 
temperatures under a field of 100 Oe.}}
\label{fig:10}
\end{figure}

{It is also important to test the universal nature of our above-mentioned picture of negative magnetization phenomenon when other high-magnetic-moment, rare-earth  ions are used in place of Gd ions in perovskite manganites.
To this end Dy$_{0.6}$Ca$_{0.4}$MnO$_3$ has been studied.
This compound has been selected because, in the bulk form, it shows a slight ZFC-FC M(T) crossing but no negative magnetization in the FC M(T) curve whereas negative magnetization is observed at  lower concentrations (i.e., $x$ = 0.2, 0.25, and 0.3) \cite{5}.}

{Bulk form and variously-sized nanoparticle  forms  of Dy$_{0.6}$Ca$_{0.4}$MnO$_3$ have been prepared and magnetization studies have been carried out. Dy$_{0.6}$Ca$_{0.4}$MnO$_3$\,-\,bulk and Dy$_{0.6}$Ca$_{0.4}$MnO$_3$\,-\,900 nanoparticle form show small bifurcation 
with small hump indicating ferromagnetic ordering. On further decreasing the particle size, Dy$_{0.6}$Ca$_{0.4}$MnO$_3$\,-\,700 nanoparticle form showed ZFC-FC crossing and negative magnetization in the FC curve (see Fig. \ref{fig:10}). Thus, the emergence of negative magnetization in the 
nano form of  
$R_{1-x}D_x$MnO$_3$ perovskite manganites (with $R^{3+}$ being a high-magnetic-moment rare-earth ion)
by controlling the system size seems to be a general phenomenon in manganites which needs extensive investigation.
}

{We would like to point out that, although  Sm$_{1-x}$Sr$_x$MnO$_3$ (for $x \le 0.05$) shows negative magnetiztion \cite{ssmo},
it belongs to a different type of systems where 
a disordered paramagnetic sublattice is
coupled with an ordered canted-antiferromagnetic  sublattice via negative-exchange interaction. We aim to study such systems in future.}

In future, we also would like to examine the deeper connection between multiferroicity in $R{\rm MnO_3}$ compounds ($R{\rm =Gd,~Tb,~ and ~Dy}$) \cite{kimura2,tokura1,kimura3,cheong2,arima} and negative
magnetization that results when $R{\rm MnO_3}$ systems are doped by  divalent alkaline elements Sr, Ca, etc.  

Lastly,  it would be useful to develop materials with temperature-induced
magnetization reversal in the vicinity of room temperature as they have potential for application in
magnetic memories such as thermally
assisted magnetic random access memory.
 }

\section{Acknowledgment}
One of the authors (S.Y.) would like to acknowledge useful discussions with K. Pradhan, I. Das, and G. Baskaran.
Papri Dasgupta is thankful to the Department of Science and Technology
(DST), India for financially supporting this study through Women Scientist
Project (Ref. No. SR/WOS-A/PM-24/2017).
\clearpage
\appendix
\begin{widetext} 
\section{Hysterisis loops}
The hysteresis curves, depicted in Fig. \ref{fig:9}, reveal the expected features that as temperature decreases magnetization
   increases both in bulk and nano samples; at lower temperatures, the curves tend towards saturation faster.
   The unusual and interesting trends of the remnant magnetizations (obtained from the hysteresis curves
   at various temperatures, dopings, and sample sizes) have been discussed
   in the main text.
   
 \vspace{1.0cm}  
  
\begin{figure}
[htb!]
\includegraphics[width=1.2\linewidth,angle=0]{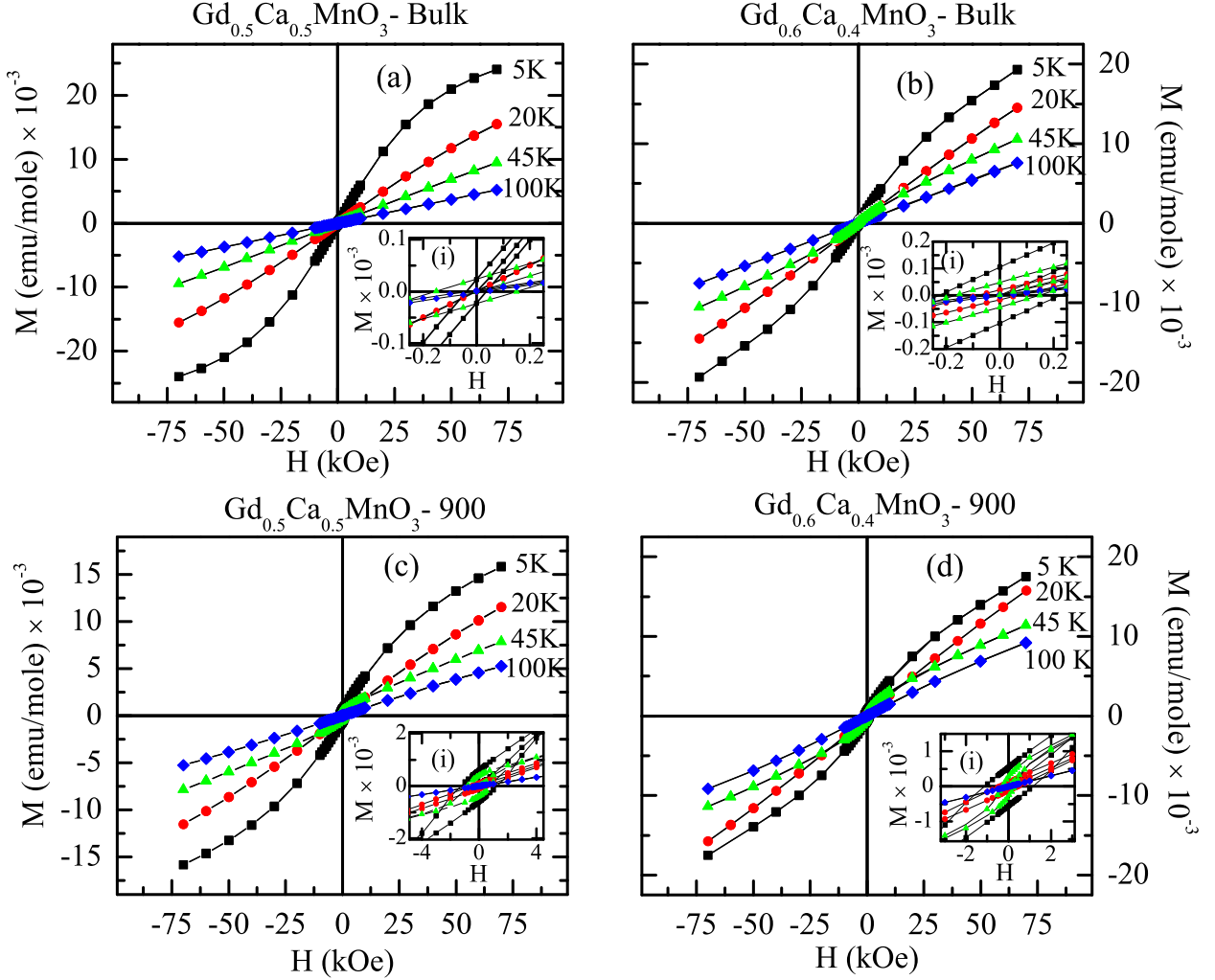} 
\vspace{-2.5cm}
\caption{(colour online) {Magnetization as a function of field at different temperatures in bulk and nano-form samples of Gd$_{1-x}$Ca$_x$MnO$_3$ 
at dopings $x=0.4$ and $0.5$. 
Inset shows enlarged version near the origin.
}}
\label{fig:9}
\end{figure}
 
\end{widetext}

\vspace{1.0cm}

\section{Bulk form magnetization at stronger magnetic fields and various dopings}   

In Fig. \ref{fig:6}, we depict that, as expected, the magnetization increases at higher magnetic fields for both the FC and the ZFC
   cases in  bulk  forms of GCMO samples;  the curves shift upwards. Furthermore, the ZFC-FC bifurcation reduces with increasing fields. 
   
   \begin{figure}
[htb!]
\includegraphics [width=0.9\linewidth,angle=0]
{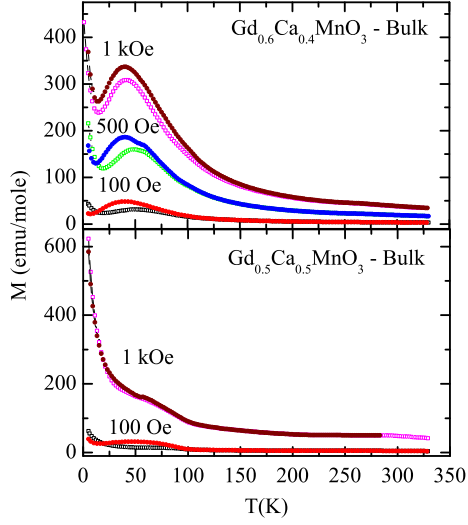}
\vspace{-1.7cm}
\caption{(colour online) {Temperature dependent magnetization curves under different applied magnetic fields in bulk Gd$_{1-x}$Ca$_x$MnO$_3$ 
{(at $x = 0.4$ and 0.5)}. At any field, the FC curves have a higher peak.}}
\label{fig:6}
\end{figure}

\section{Transition from negative to positive magnetization due to increasing magnetic field  in nano-form GCMO}  
In Fig. \ref{fig:8} we demonstrate that, as the particle size increases in nano-form Gd$_{0.5}$Ca$_{0.5}$MnO$_3$, 
a smaller magnetic field is needed to produce a positive
magnetization at lower temperatures.
\begin{figure}
[htb!]
\vspace{-1cm}
\includegraphics[width=1.9\linewidth,angle=0]{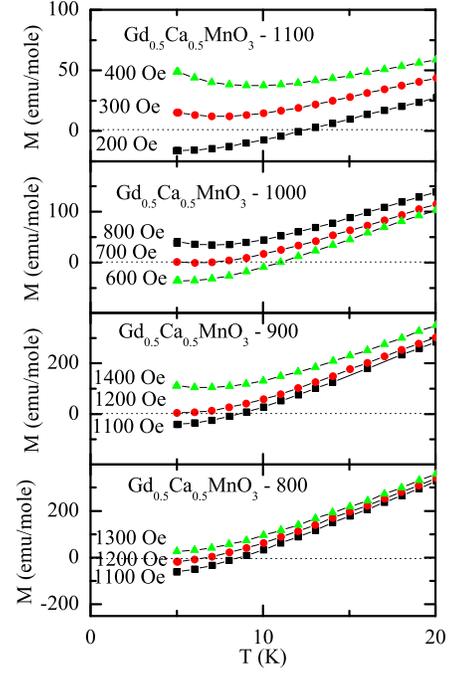}
\vspace{-3cm}
\caption{(colour online) {Temperature variation of field-cooled magnetization curves M(T) 
under different applied magnetic fields in Gd$_{0.5}$Ca$_{0.5}$MnO$_3$\,-\,800, Gd$_{0.5}$Ca$_{0.5}$MnO$_3$\,-\,900, 
Gd$_{0.5}$Ca$_{0.5}$MnO$_3$\,-\,1000 and Gd$_{0.5}$Ca$_{0.5}$MnO$_3$\,-\,1100 nanoparticle forms.}}
\label{fig:8} 
\end{figure}


\clearpage





\begin{thebibliography}{99}
\bibitem{cnr}
{\em Colossal Magnetoresistance, Charge Ordering, and Related
Properties of Manganese Oxides}, edited by C.N.R. Rao and
B. Raveau (World Scientific, Singapore, 1998).
\bibitem{khomskii0}
D. I. Khomskii, Physica Scripta {\bf 72}, CC8 (2005).
\bibitem{9} E. Dagotta, T. Hotta, and A. Moreo, Physics Reports {\bf 344}, 1 (2001).
\bibitem{hotta}
T. Hotta, Rep. Prog. Phys. {\bf 69}, 2061 (2006).
\bibitem{tokura}
Y. Tokura, Rep. Prog. Phys. {\bf 69}, 797 (2006).
\bibitem{cheong}
See K.H. Kim, M. Uehara, V. Kiryukhin and S.-W. Cheong, in 
{\em Colossal Magnetoresistive Manganites}, edited by T. Chatterji, (Kluwer Academic, Dordrecht, 2004).
\bibitem{raveau}
 C. Martin, A. Maignan, M. Hervieu, and B. Raveau,
 Phys. Rev. B {\bf 60}, 12191 (1999).
\bibitem{pauthenet} 
R. Pauthenet and C. Veyret, J. Phys. (Paris) {\bf 31}, 65 (1970).
\bibitem{kimura}
T. Kimura, S. Ishihara, H. Shintani, T. Arima, K. T. Takahashi, K. Ishizaka, and Y. Tokura,
Phys. Rev. B {\bf 68}, 060403(R) (2003).
\bibitem{shannon}
R. D. Shannon, Acta Crystallogr. Sec. A {\bf 32}, 751 (1976).
\bibitem{yusuf}
A. Kumar and S. M. Yusuf, Physics Reports {\bf 556}, 1 (2015).
\bibitem{biswas}
S. Biswas and S. Pal, Rev. Adv. Mater. Sci. {\bf 53}, 206 (2018).
\bibitem{kimura2}
T. Kimura, T. Goto, H. Shintani, K. Ishizaka, T. Arima, and Y.
Tokura, Nature (London) {\bf 426}, 55 (2003). 
\bibitem{tokura1}
T. Goto, T. Kimura, G. Lawes, A. P. Ramirez, and Y. Tokura,
Phys. Rev. Lett. {\bf 92}, 257201 (2004).
\bibitem{kimura3}
T. Kimura, G. Lawes, T. Goto, Y. Tokura, and A. P. Ramirez,
Phys. Rev. B {\bf 71}, 224425 (2005).
\bibitem{cheong2}
M. Kenzelmann, A. B. Harris, S. Jonas, C. Broholm, J. Schefer,
S. B. Kim, C. L. Zhang, S.-W. Cheong, O. P. Vajk, and J. W.
Lynn, Phys. Rev. Lett. {\bf 95}, 087206 (2005).
\bibitem{arima}
T. Arima, T. Goto, Y. Yamasaki, S. Miyasaka, K. Ishii, M.
Tsubota, T. Inami, Y. Murakami, and Y. Tokura, Phys. Rev. B
{\bf 72}, 100102(R) (2005).

\bibitem{4} O. Pena, M. Bahout, K. Ghanimi, P. Duran, D. Gutierrez, and C. Moure, J. Mater. Chem. {\bf 12}, 2480 (2002).
\bibitem{5} M. M. Bahout, O. Pena, D. Gutierrez, P. Duran, and C. Moure, Solid State Communications {\bf 122}, 561 (2002).
\bibitem{snyder}
G. J. Snyder, C. H. Booth, F. Bridges, R. Hiskes, S. DiCarolis,
M. R. Beasley and T. H. Geballe, Phys. Rev. B {\bf 55}, 6453 (1997).
\bibitem{ma} Y. Ma, M. Guilloux-Viry, P. Barahona, O. Pena, and C. Moure,  Appl. Phys. Lett. {\bf 86},  062506 (2005).
\bibitem{yingnan} Z. Yingnan, L. Junjia, Z. Ziqing, L. Fuyang,
Z. Xudong and L. Xiaoyang, Chem. Res. Chin. Univ. {\bf 31},  699 (2015).
\bibitem{nagaraja} B. S. Nagaraja, A. Rao, P. D. Babu and G.S.
Okram, Physica B {\bf 479}, 10 (2015).
\bibitem{jin} Y. Jin, X. P. Cui, J. A. Cheng, S. X. Cao, W. Ren and J. C. Zhang, Appl. Phys. Lett. {\bf 107},  072907 (2015).
\bibitem{elizabeth} H. Nhalil and S. Elizabeth, J. Supercond. Nov. Magn. {\bf 30},  1681 (2017).
\bibitem{7} S. M. Yusuf, A. Kumar, and J. V. Yakhmi, Appl. Phys. Lett. {\bf 95}, 182506 (2009).
\bibitem{8} A. Poddar, S. Das, and B. Chattopadhyay, J. Appl. Phys. {\bf 95}, 6261 (2004).
\bibitem{10} A. Beiranvand, J. Tikkanen, H. Huhtinen, and P. Paturi, Journal of Alloys and Compounds {\bf 720}, 126 (2017).
\bibitem{11} C. Moure and O. Pena, J. Mag. Mag. Materials {\bf 337-338},  1 (2013).
\bibitem{paturi2} A. Beiranvand, J. Tikkanen, H. Huhtinen, and P. Paturi, J. Mag. Mag. Materials {\bf 469},  253 (2019).
\bibitem{11a} J.Rodriguez-Carvajal, Physica B {\bf 192}, 55 (1993).

\bibitem{gennes}
P. -G. de Gennes, Phys. Rev. {\bf 118}, 141 (1960).
\bibitem{Izyumov}
Y. A. Izyumov and Y. N. Skryabin, Phys.-Usp. {\bf 44}, 109 (2001).
\bibitem{hwang}
H. Y. Hwang, S.-W. Cheong, P. G. Radaelli, M. Marezio, and
B. Batlogg, Phys. Rev. Lett. {\bf 75}, 914 (1995).
\bibitem{ME}
S. Paul, R. Pankaj, S. Yarlagadda, P.  Majumdar, and P. B. Littlewood,
Phys. Rev. B {\bf 96}, 195130 (2017).
\bibitem{sdadsy1}
 S. Datta, A. Das, and S. Yarlagadda, Phys.
Rev. B {\bf 71}, 235118 (2005).
\bibitem{sp}
S. Paul and S. Yarlagadda,
Phys. Rev. B {\bf 103}, 035140 (2021). 
\bibitem{anderson}
P. W. Anderson, Phys. Rev. {\bf 115},  2 (1959).


\bibitem{13} K. Das, T. Paramanik and I. Das, J. Mag. Mag. Mater. {\bf 374}, 707 (2015).
\bibitem{ssmo} V. Yu. Ivanov, A. A. Mukhin, A. S. Prokhorov, and A. M. Balbashov, Phys. Stat. Sol. (b) {\bf 236}, 
445 (2003).


\end{thebibliography}
\end{document}